\documentclass[12pt]{article}
\usepackage[dvips]{graphicx}

\newcommand{\gsim} {\buildrel > \over {_\sim}}
\newcommand{\VEV}[1]{\left\langle{#1}\right\rangle}
\renewcommand{\bar}[1]{\overline{#1}}

\begin {document}
\begin{flushright}
{\small
SLAC--PUB--10276\\
26 March 2004\\}
\end{flushright}

\vfill

\begin{center}
{{\bf\LARGE The Classification of Universes}\footnote{Work
supported in part by Department of Energy contract
DE--AC03--76SF00515.}}

\bigskip
{\it James D. Bjorken \\
Stanford Linear Accelerator Center \\
Stanford University, Stanford, California 94309 \\
E-mail:  bjorken@slac.stanford.edu}
\medskip
\end{center}

\bigskip

\begin{center}
Abstract
\end{center}

We define a universe as the contents of a spacetime box with
comoving walls, large enough to contain essentially all phenomena
that can be conceivably measured. The initial time is taken as the
epoch when the lowest CMB modes undergo horizon crossing, and the
final time taken when the wavelengths of CMB photons are
comparable with the Hubble scale, {\em i.e.} with the nominal size
of the universe. This allows the definition of a local ensemble of
similarly constructed universes, using only modest extrapolations
of the observed behavior of the cosmos. We then assume that
further out in spacetime, similar universes can be constructed but
containing different standard model parameters. Within this
multiverse ensemble, it is assumed that the standard model parameters are
strongly correlated with size, {\em i.e.} with the value of the
inverse Hubble parameter at the final time, in a manner as
previously suggested. This allows an estimate of the range of
sizes which allow life as we know it, and invites a speculation
regarding the most natural distribution of sizes. If small sizes
are favored, this in turn allows some understanding of the
hierarchy problems of particle physics. Subsequent sections of the
paper explore other possible implications. In all cases, the
approach is as bottoms up and as phenomenological as possible, and
suggests that theories of the multiverse so constructed may in fact lay some
claim of being scientific.

\vfill \newpage

\section{Introduction}

The idea that the universe we see around us is a member of an
ensemble of universes (the multiverse), the remainder of which are
beyond our view, is an old one, But it is one under active
investigation nowadays. Much of the contemporary motivation comes
from rather grandiose theoretical conceptualizations. These
include, but are not limited to, string theory ideology, eternal
inflation, and even Everett's many-worlds interpretation of
quantum mechanics~\cite{ref:a}.

But another motivation comes from the great progress being made in
the experimental investigation of the structure and early history
of our own universe, as well as the successful establishment of
the standard model of particle physics. While the accumulation of
these data is no doubt the best stimulant, it nevertheless has led
to a mixing of direct information from the experiments with highly
speculative material, remote from experimental test. Indeed, there
is plenty of skepticism extant as to whether the consideration of
universes which are causally disconnected from our own is science
at all.

The purpose of this note is to make the study of ensembles of
universes as data-driven as possible. This will entail, first, a
specific definition of a universe as the contents of a comoving
spacetime box, large enough to contain essentially everything we
can expect to be able to measure (based on the present consensus
picture of cosmology and particle physics), but still small
compared to the extent of the total spacetime domain envisaged by
the majority of cosmologists and other theorists. This allows a
credible way of constructing a local ensemble of universes very
similar to our own.  One simply generalizes the construction of
our universe to other spacetime domains causally disconnected from
our own, yet near enough to allow reasonable extrapolation of
properties of our universe to the remainder of the local ensemble.

We envisage this construction of our universe in analogy to the
establishment of a homestead in Kansas by a pioneer family. They
would see their spatial environment, as we do see our universe, as
flatter than a pancake~\cite{ref:c}, all the way to their property
line. And it would be not unreasonable---and in fact even
correct---for them to envisage other similar homesteads beyond
their horizon. Were the family to contain a theorist, he or she
would in fact probably erroneously conclude that Kansas, and hence
the number of such homesteads, in fact was infinite in
extent~\cite{ref:d}. However, far beyond Kansas will be found
other pieces of fertile terrain, {\em e.g.} in the Central Valley
of California, which also are flatter than a pancake and which
bear other similarities to, as well as differences from, the
Kansas homesteads.

The local, ``Kansas" ensemble of universes we shall consider will
be assumed to possess a spatially flat FRW metric,  with similar
initial cosmological boundary conditions. This ensemble is already
phenomenologically of some use, because it can be considered as
the ensemble implicitly used by inflationary cosmologists in
making quantum averages of the inflationary perturbation spectra.
But more interesting, and the focus of this note, are the
analogues of the Central Valley homesteads. We take these to be
similar to our universe in the sense of again being described by a
spatially flat FRW metric, with perhaps some also being fertile,
{\em i.e.} capable of supporting life as we know it. The features
that we assume can vary from universe to universe will include the
parameters of the standard model and of cosmology.

One can generalize much further than this, at the cost of
increasing the level of speculation~\cite{ref:e}. We will stop at
this point, however. The reasons are, first, that obviously this
level is already quite speculative, and second, that the
data-driven (as well as theoretical!) motivations for considering
ensembles of universes can already be expressed with specificity
at this level of speculation. The main data-driven consideration
is that the values of many standard model parameters seem to be
fine-tuned in a way to allow our existence in the universe. The
existence of an ensemble of universes with a variety of
standard-model parameters allows an ``anthropic" interpretation of
this property of standard-model data~\cite{ref:a,ref:f}. And the
main theoretical consideration comes from string theory, where a
variety of string vacuua, with different standard-model
properties, may easily be envisaged to occur in various causally
disconnected regions of spacetime~\cite{ref:g}.

These issues can all begin to be addressed in the context of the
subensemble of universes that we construct, namely those that
differ in a rather minimal way from our own. It is already of
great interest to try to delineate the existence and properties of
these ``nearest-neighbor" universes. And the possibility of
erroneous speculation, while still huge, is clearly going to be
less than dealing (in the absence of data) with a more broadly
generalized ensemble of universes possessing properties vastly
different from our own.

Another analogy to the ensemble of universes is the perhaps
familiar one of the planets in our own universe. These are already
known to come in many varieties, but an especially interesting
subset of planets consists of those containing life as we know it.
This ``nearest-neighbor" ensemble might consist of only Planet
Earth, or it may contain a large number~\cite{ref:h}. The question
of which alternative is correct is not only irresistible to ask,
but is also undeniably a scientific one. In order to address it in
a systematic way, one needs to parametrize properties of planets
in such a way that one can look at the distribution function of
those parameters and find the size and location of the habitable
island in parameter space. The output of such a program could be
an estimate or bound on the number of planets which can in
principle support life as we know it. This is in the case of
planets a very daunting task, one that will need to be data-driven
to make progress~\cite{ref:i}. If universes as we have defined
them have a similarly complex parametrization, it will be
extremely difficult to make progress, simply due to lack of data.
It is a serious criticism by those who remain skeptical of the
utility of multiverse ideology that this problem presents an
insuperable barrier to scientific consideration of this line of
thinking.

The use of ``anthropic" reasoning when dealing with ensembles of
universes only exacerbates this problem. If the distribution
function of standard model parameters, taken over the multiverse
ensemble, is broad in each of the parameters, and if there is
little correlation between the probability of finding the value of
one parameter and that of another, then the program of finding
naturalistic explanations for the values of standard model
parameters essentially grinds to a halt. The replacement is a
statement, unsupported by data, that they are determined by
historical accident. To this author, such an option seems not much
more scientific in nature than interpreting the values of the
standard model parameters via ``intelligent design".

However, we cannot ourselves impose a decision on how things work
at this level. If the above option is true, then it may be that
the scientific method will never have the power to find out the
answers to the Big Questions. But, it is not clear that the
problem of characterizing the ensemble of universes is in fact as
difficult as characterizing the ensemble of habitable planets. As
long as there is an outside chance that the problem is simpler and
more tractable, there is no reason not to pursue that possibility.
This optimistic position will in fact be our working assumption.
We shall assume that many of the standard model parameters are
strongly correlated with each other, thereby simplifying the
nature of the aforementioned distribution function. While
anthropic reasoning will be utilized, the amount will be limited
by the above simplistic assumption of the existence of
correlations. And the test of whether the assumption makes sense
will be whether it delivers insight into any of the unsolved
problems facing particle physics and cosmology. We shall argue on
the basis of the results that the answer to this is positive. In
particular we shall obtain a partial understanding of the nature
of the hierarchy problems of particle physics: why the ratios of
many of the standard model parameters are such large numbers. We
do not claim that the hierarchy problems are thereby solved, but
rather that they are recast in a different form, one which in turn
appears to be able to be attacked by scientific methods.

In the next section, we discuss in detail our definition of a
universe, and in Section 3 provide a concise description of its
major properties. In Section 4 we review the nature of our assumed
correlations of standard model parameters , emphasizing their
simplicity and cogency, as well as the consequence that the
habitable range of sizes is limited to within a factor two of the
size of our universe. In Section 5, we discuss the anthropic
implications of this scenario, in particular that if the
distribution in sizes is peaked at small sizes, there results some
understanding of the hierarchy problems of particle physics.

The above material contains the main messages of this paper. The
remaining sections are spinoff topics even more speculative.
Section 6 explores a possible connection between
Bekenstein-Hawking horizon entropy and the vacuum structure of
QCD. Section 7 explores whether the inflationary phase of our
universe might have its own set of standard model parameters,
correlated with the very large value of the Hubble parameter (dark
energy) during that epoch. Section 8 raises the question of
whether the starting time we have used in our construction of the
universe may contain physics, and represent the actual onset of
the inflationary epoch. Section 9 is devoted to concluding
comments.

\section{The Standard Universe}

We define our universe in rough terms as ``what we and our
descendants can observe in principle, plus a little more". To be
more specific, we need some mild assumptions, mild at least in
comparison to others which will follow. They include:

1) The observable universe obeys on large scales the cosmological
principle, and is described by the FRW metric.

2) The observable universe is spatially flat.

3) In the distant future, the universe will reinflate, {\em i.e.}
be characterized by a nonvanishing cosmological constant.  We also
assume that quintessence is either absent or unimportant on the
time scales which we consider.

4) The broad history of our universe is well-described by the
``concordance model":  an inflationary epoch, followed by
``reheating", the radiation-dominated big bang, and subsequent
events, as described in  standard texts and
references~\cite{ref:j}.

5) Our preferred inflationary scenarios lie in the category of
``hybrid" models~\cite{ref:k}. In particular we assume that the
evolution during the inflationary epoch we consider is
``quasi-deSitter", with, to good accuracy, a nearly
time-independent Hubble parameter.

With these assumptions, we may now map out the spacetime region of
the observable portion of the universe. To do this we start with
the form of the FRW metric
\begin{equation}
ds^2 = dt^2 - a^2(t)[d\theta^2_x + d\theta^2_y + d\theta^2_z] \ .
\label{eq:1}
\end{equation}
All the information is codified in the time dependence of the
scale factor $a(t)$. Its dynamics is determined in a simple way by
the Einstein equations, which we write down later. But before
discussing that we introduce another useful set of coordinates,
the ``conformally flat coordinates". The time $t$ is traded in for
``conformal time" $\eta$, defined as
\begin{equation}
\eta = - \int^\infty_t \ \frac{dt}{a(t)}
\qquad d\eta = \frac{dt}{a(t)} \ .
\label{eq:2}
\end{equation}
Therefore the line element in Eq. (\ref{eq:1}) becomes
\begin{equation}
ds^2 = a^2(t) [d\eta^2 - d\theta^2_x - d\theta^2_y - d\theta^2_z]
\ . \label{eq:3}
\end{equation}
The great usefulness of this choice is the clarity in seeing the
causal connection between different spacetime events, since the
trajectories of light rays in these coordinates are straight lines
at 45 degrees to the vertical, just as in Minkowski spacetime.
Therefore the portion of the FRW spacetime which we can see is a
past light cone emanating from our worldline, which evidently is
along the (conformal) time axis.

\begin{figure}
\begin{center}
\includegraphics[width=0.8\textwidth]{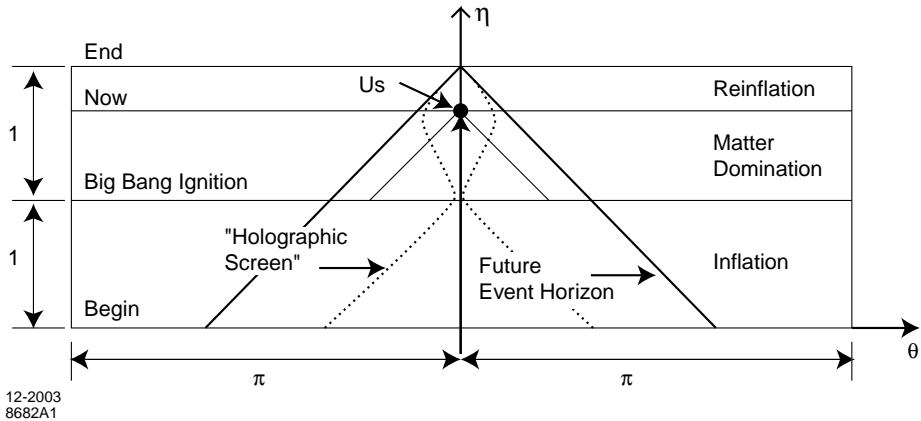}
\end{center}
\caption{``Conformal" spacetime.  The surface marked ``holographic
screen" is simply defined by $r = a(\eta)|\, \vec{\!\theta}\,| =
H^{-1}(\eta)$.} \label{fig:a}
\end{figure}

Important is the fact, implied in Eq.~(\ref{eq:2}), that infinite
FRW time maps into a finite, maximal conformal time. This occurs
because of the existence of dark energy. The big-bang power-law
growth of $a(t)$ morphs into an exponential growth once the dark
energy becomes the dominant source term in the Einstein equations,
thereby providing convergence for the integral in
Eq.~(\ref{eq:2}). At present, we already are able to see a large
fraction of the spacetime that any distant descendants of us will
be able to see. The past lightcone emanating from the $t = \infty$
(or $\eta = 0$) point on our worldline, as shown in Fig.
\ref{fig:a}, provides a natural boundary between what is clearly a
part of our own universe (as defined in the beginning of this
section) and what is arguably beyond. This boundary surface is
called by Hawking and Ellis~\cite{ref:xx} the ``future event
horizon".

While we might entertain defining the future event horizon as the
boundary of our universe, we shall not do so. Instead, we shall
define it as the interior of a comoving box. This is defined at
any particular $t$ or $\eta$ as an ordinary box of physical size
$L= 2\pi a(t)$ on a side, with any point on that box moving along
the $t$ or $\eta$ axis. This means that at sufficiently early
times the box will lie inside the future event horizon. However
our experimental access to extremely early times is both in
practice and in principle very limited. So we shall also limit the
spacetime of our universe to be in the future of some initial FRW
time. This time is defined, roughly, as the time when, say, the
dipole component of the cosmic microwave background undergoes
``horizon crossing", although the exact choice is negotiable.
However, at this initial time, we shall require that the size of
our comoving box be larger than the size of the future event
horizon.

Likewise, we shall terminate the universe at a future time, of
about one trillion years. This is a landmark time; the cosmic
microwave background radiation is redshifted so much that the
wavelength of the photons becomes comparable to the characteristic
size of the universe. Thereafter, all the galaxies except those in
our local cluster will have receded from view and the visible part
of the universe will be filled with Hawking radiation of a
wavelength typical of the size scale of the universe. The
``classical" phase of the accelerated, dark-energy driven,
deSitter expansion of the universe will terminate and be
supplanted by the quantum deSitter expansion, about which there is
much more theoretical uncertainty~\cite{ref:xy} and angst. So it
seems like a good occasion for drawing the line between credible
extrapolation and less credible speculations.

The net effect of these constructions is to define a spacetime
box, within which lies essentially all the phenomenology that we
can presently perceive and access, and beyond which exists a huge
amount of theoretical and philosophical ideas, unsupported by
data. The boundaries of the box are artificial, but it is unlikely
that boundary effects exist that affect the phenomenology within.
And the choice of a comoving box instead of some other boundary is
strongly motivated by practical considerations. The graviton and
inflaton modes which generate the cosmic microwave background
texture are most simply described in terms of a mode expansion
utilizing comoving-box boundary conditions, which we for
simplicity take as periodic.

The contents of the spacetime box which we have constructed is by
definition what we shall call a universe. Its parametrization, in
particular the definition of the initial and final time surfaces,
as well as choice of spatial size, will be discussed below in more
detail. But before moving on, it should be emphasized that this
definition of universe finesses many interesting questions, such
as the origin of time, how and when the inflationary epoch begins,
the ultimate fate of the universe, and the extent in space over
which the FRW description remains valid. It is not that these
questions are not important. But they are by definition ``outside
the box", and therefore relatively remote from the issues that are
accessible phenomenologically.

\section{Describing the Universe}

\begin{figure}
\begin{center}
\includegraphics[width=0.8\textwidth]{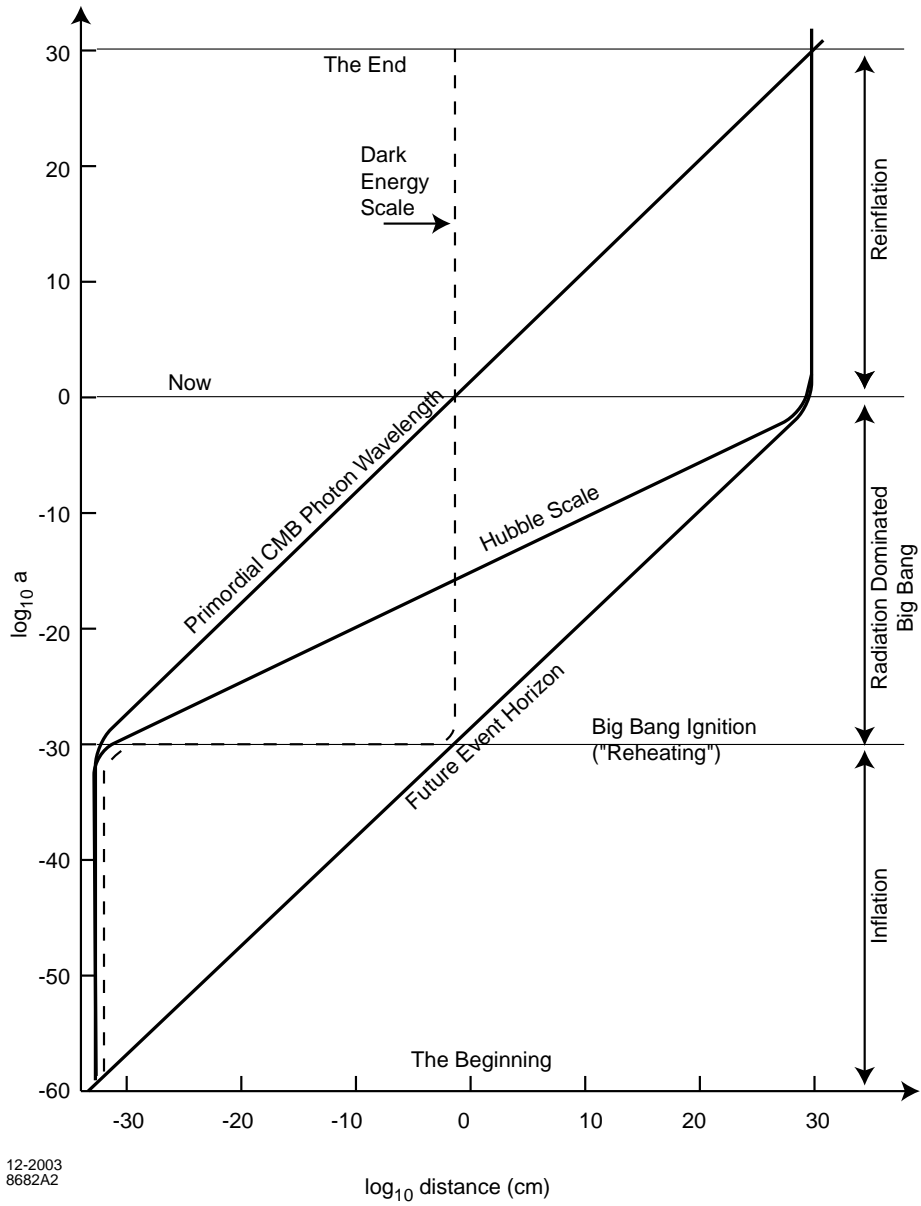}
\end{center}
\caption{The ``universe for dummies".} \label{fig:b}
\end{figure}

The gross properties of our universe are encoded in the properties
of the FRW scale factor $a(t)$, and in the dependence of other
parameters of the universe on that quantity. We shall first look
at these properties in a highly simplified limit, which we call
"the universe for dummies". In this limit, we disregard the dark
and hadronic matter content of the universe, and also disregard
the difference between the scale of inflationary dark energy and
the Planck scale. This involves setting various factors of up to
$10^5$ to unity, but will involve keeping remaining factors of
$10^{30}$ intact. The result is shown in Fig. \ref{fig:b}, where
on a log-log scale various quantities with dimension of distance
are plotted against the scale factor, which in turn can be
regarded as a time variable. These quantities include (1) the
Hubble parameter
\begin{equation}
H = \frac{\dot{a}}{a} = \frac{d}{dt} \, (\ell n\, a)
\label{eq:4}
\end{equation}
(2) the magnitude of the characteristic dark energy distance scale
$\mu^{-1}$ in the inflationary and present-day epochs.  Here
$\mu^4$ is the value of the dark energy density term in the
standard-model Lagrangian density, related to the Hubble parameter
of a dark-energy dominated expansion by
\begin{equation}
 \mu^4 = \frac{3}{8\pi} H^2 M_{pl}^2 \ ,
\label{eq:5}
\end{equation}
(3) the physical radius of the future event horizon $E(t)$, as
discussed in the previous section, defined as
\begin{equation}
  E(t) = a(t) |\eta|
\label{eq:6}
\end{equation}
and (4) the physical wavelength of the cosmic microwave background
radiation.

We see from Fig. \ref{fig:b} that the basic architecture is
characterized by three epochs of cosmic expansion, each
characterized by an increase in the FRW scale factor $a(t)$ by a
factor of about $10^{30}$. Likewise the ratio of cosmological
scale to dark energy scale is again a factor $10^{30}$, as is the
ratio of dark energy scale to the Planck scale. Evidently it is
the cosmological constant which establishes this overall
architecture.

We infer from this feature that the parameter most responsible for
the gross features of our universe is in fact the cosmological
constant. We consequently identify the intrinsic size of a
universe with the value of its inverse Hubble parameter at late
times, related to the cosmological constant by Eq. (\ref{eq:5}).
We consider this  definition of size to be an important
characterization, perhaps the most important one, possessed by our
universe and by other members of the ensemble of universes that we
shall consider.\footnote{Note:  This definition of size is NOT the
same as the scale size of our own universe, proportional to
$a(t)$.  We assume NO standard-model parameter dependence with
time $t$ or scale factor $a(t)$ in our own universe.  (See
however, Section 7 for a caveat.)}

\begin{figure}
\begin{center}
\includegraphics[width=0.8\textwidth]{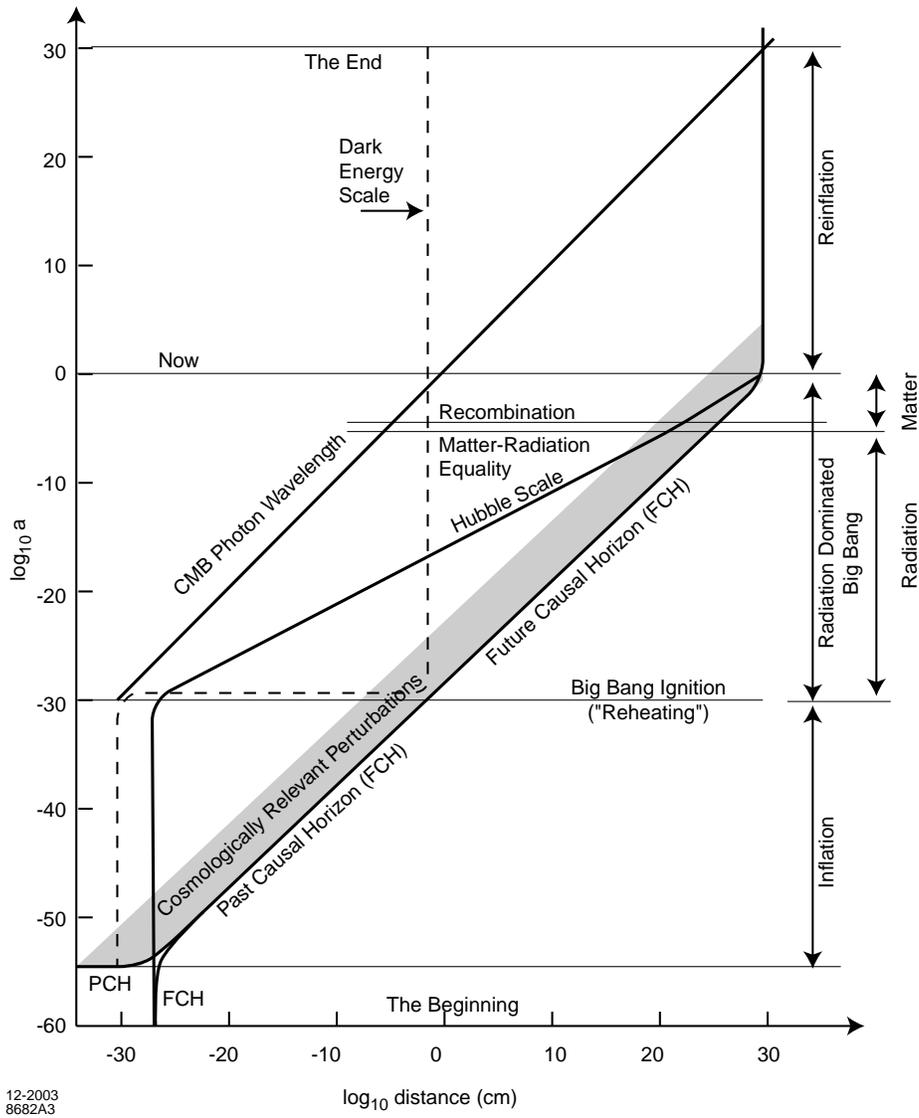}
\end{center}
\caption{The real universe.}
\label{fig:c}
\end{figure}

We also see from  Fig. \ref{fig:b} that the initial time of our
universe-in-a-box is naturally taken as when the value of the
future event horizon is comparable to the value of the inverse of
the Hubble parameter, and that the final time is naturally taken
when the CMB wavelength  is comparable to the inverse Hubble
parameter.

A more realistic version of our universe is shown in Fig.
\ref{fig:c}. We have here included the modification of the
evolution of the Hubble parameter due to the presence of dark
matter. We also have parametrized the inflationary epoch more
realistically, taking the inflationary dark energy scale to be of
order the grand unification scale of about $10^{16}$ GeV, in rough
concordance with the consensus value used by perhaps a majority of
inflation theorists. This diminishes the Hubble parameter relative
to the Planck scale by a very significant factor of about $10^{-5}
- 10^{-6}$. We also show in Fig. \ref{fig:c} the size of the
inflationary perturbations relevant to phenomenology. One sees the
``horizon-crossings", namely the times at which the physical
wavelengths of the perturbations are comparable to the inverse
Hubble scale. It is clear from Fig. \ref{fig:c} that, as promised
above, the initial time assumed in the construction of the
spacetime box is comparable to the time when the lowest CMB
multipoles ``cross the horizon".  In fact it is more precisely
defined by the requirement that the conformal time interval
$\Delta\eta$ during inflation equals the $\Delta\eta$ from
big-bang ignition to the final time we consider (essentially
infinite FRW time $t$).

Our favorite choice for the size of the spacetime box is shown in
Fig. \ref{fig:b}. The spatial dimension is $2 \pi$ larger than the
duration of conformal time $\eta$ from reheating to the end, or
from the beginning of inflation to the reheating time. This makes
the values $k$ of the plane-wave mode expansion for the box in one
to-one correspondence with the values of $\ell$ that our
descendants at late times would use in analyzing the primordial
CMB fluctuations. More importantly, this choice of spatial size
should be large enough to make spurious boundary effects
negligible phenomenologically, but without a great deal to
spare~\cite{ref:k}.

A major advantage of the description of the evolution of the
universe of Fig. \ref{fig:c} is that all the major epochs and
distance or energy scales are readily visible in the plot; no
features are squished into a tiny region. This is not the case in
Fig. \ref{fig:a}, where almost everything interesting is squished
into the central line labelled ``Big Bang Ignition" (reheating),
or into the tip of the future-event-horizon light cone. There are
only two epochs which really show clearly in Fig. \ref{fig:a}. One
is the present epoch, extending from a redshift of two or three in
our past, to two or three efoldings of accelerated expansion in
our future. The other epoch comprises the first few efoldings of
inflation following the initial time we have specified for our
universe. For completeness, we provide some mathematical details
which further explain the two figures and their relationship in
Appendix A.

\section{Correlations of Standard Model Parameters}

In the previous section, we argued that the most gross features of
our universe, as determined by experimental cosmology, are
characterized by the value of the asymptotic Hubble parameter,
which is determined by the value of the cosmological constant. We
assume that all the members of the ensemble of universes that we
consider are likewise characterized, so that they can be labelled
by ``size", where ``size"  is defined by the value of the inverse
Hubble parameter at late times.

We now consider the hypothesis discussed in Section 1, that other
standard model parameters are strongly correlated, and in
particular are correlated with this size parameter. This idea and
a specific proposal of the nature of the correlation has already
been put forward by us. While the approach is bottoms-up
guesswork, we believe the hypothesis has a surprisingly natural
and credible character, and therefore choose to elaborate again on
it here.

The idea proceeds in two main steps. The first has been discussed
already: either strong correlations exist or they do not. If they
do not, then there is the danger that the entire set of standard
model parameters ends up in an anthropic swamp, and predictivity
is lost. The alternative requires, to be sure, optimism. But the
optimism may be necessary to keep the discourse within, or at
least reasonably near, the domain of observational science.

\begin{figure}
\begin{center}
\includegraphics[width=0.8\textwidth]{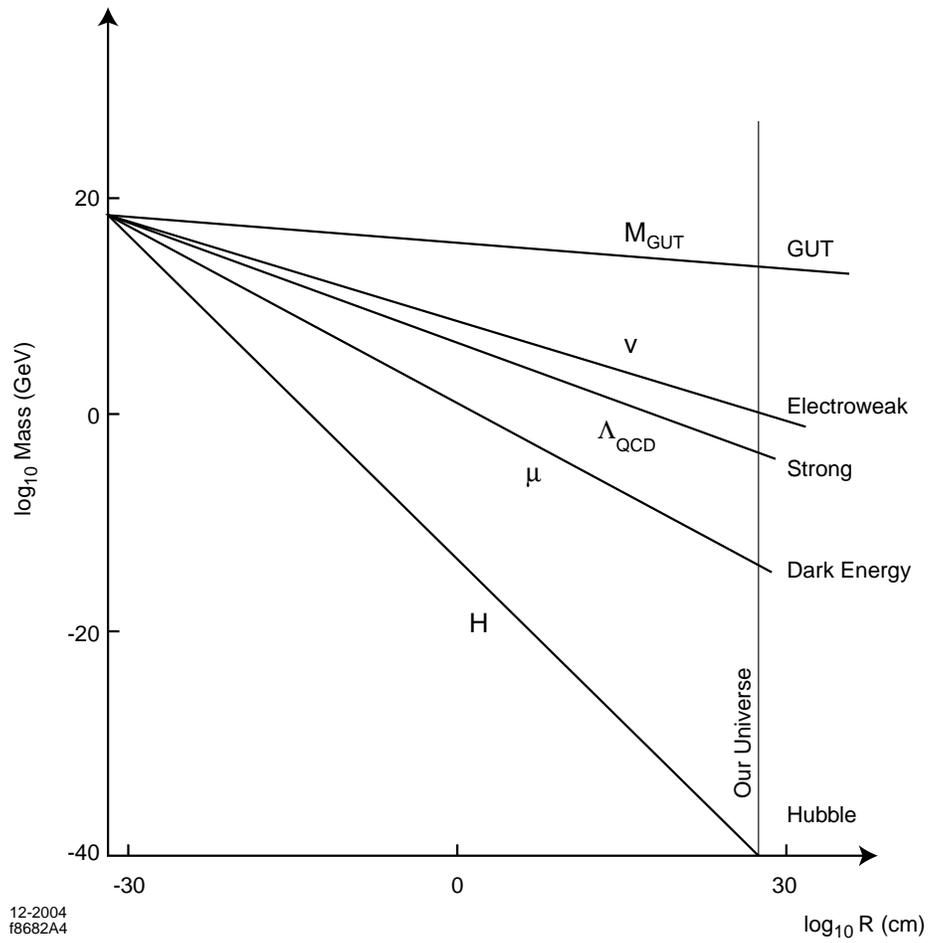}
\end{center}
\caption{Assumed size dependence of standard model vacuum
parameters.}
\label{fig:d}
\end{figure}

The second step is to assume that the standard-model vacuum
parameters are the best candidates for being strongly correlated
with the value of the dark energy, which itself is also a vacuum
parameter~\cite{ref:f}. After all, in the real theory, as opposed
to the fragmented version the standard model presents, there is
only one vacuum state for our universe. And we know by definition
that the dark-energy scale appearing in the local Lagrangian
density of the standard model does vary with the size of the
universe as we have defined it. This is shown in Fig. \ref{fig:d}
in a log-log plot, where the variation is a straight line heading
from the observed scale for our large universe to the Planck/GUT
scale for very small universes. Our assumption is that, at least
approximately, the same thing occurs for the QCD vacuum scale
$\Lambda_{\rm QCD}$ and the electroweak scale, represented by the
value $v$ of the magnitude of the Higgs condensate. The
consequences of this assumption are also shown in Fig.
\ref{fig:d}. It is as if there is renormalization-group flow of
these parameters, with one and only one fixed point in the
neighborhood of the Planck/GUT scale.

\begin{figure}
\begin{center}
\includegraphics[width=0.8\textwidth]{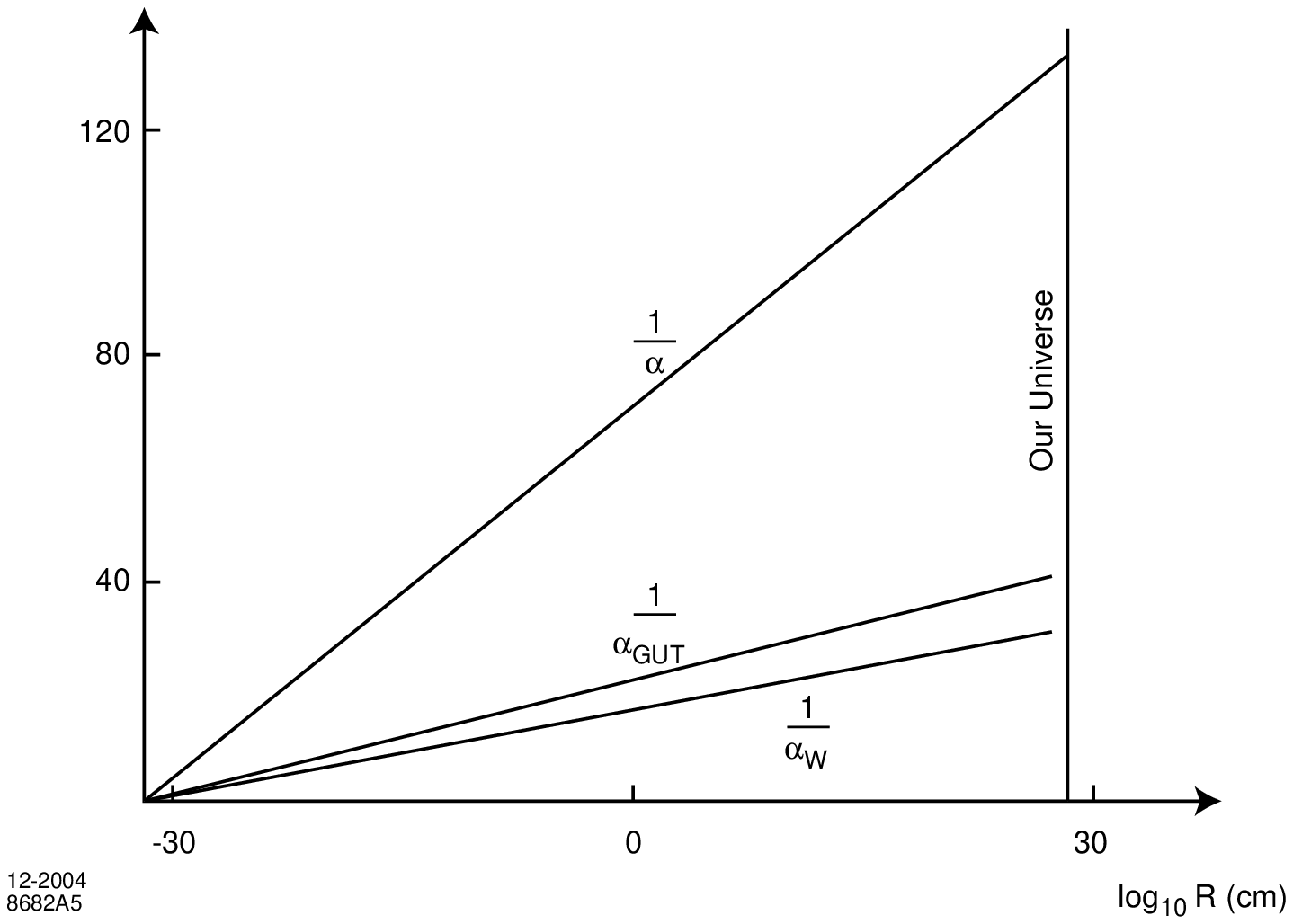}
\end{center}
\caption{Size dependence of coupling constants.}
\label{fig:e}
\end{figure}

It is our experience, based on more than a decade of playing with
this idea, that this simple hypothesis stands out as much more
credible than other versions that one might concoct. It leads in
particular to a similarly simple behavior for the gauge coupling
constants, shown in Fig. \ref{fig:e}. Because the strong coupling
constant depends upon $\Lambda_{\rm QCD}$, its behavior with the
size of the universe can be determined. Then, upon running the QCD
coupling up to the GUT scale, the size-dependence of the GUT
coupling can also be determined. This in turn allows the
size-dependence of the electroweak couplings to also be determined
by running back down from the GUT scale to the low energy scales.
It is to be emphasized that in doing all this, there are no extra
assumptions involved---only straightforward computation. And the
net result, as exhibited in Fig. \ref{fig:e}, is simple,
interesting, and at an intuitive level altogether credible: very
small universes have strong standard model interactions, while the
infinite universe would be trivial, with no interactions
remaining.  There seems to be a faint echo of the holographic
hypotheses:  existence of interactions in the bulk requires the
existence of a boundary.

The details of all this, including what to do with the remaining
mass and mixing parameters of the standard model, are explored in
the previous papers~\cite{ref:f}. It is not our intention to
retread all of this ground here, but only to highlight the main
consequences of those investigations. In brief, they are that many
properties of our universe, like the existence of chemistry and of
recognizable nuclear forces and nuclear matter, are robust and can
exist over a broad range of sizes. However, when one addresses the
issue of the range of sizes over which life as we know it can be
expected to exist, there are stronger constraints, well known to
the anthropic community~\cite{ref:n}. These constraints are
sensitive to small quantities such as the pion to proton mass
ratio, controlled by the masses of the up and down quarks,
quantities which are in the several-MeV mass range. The size
dependence of such quantities is harder to infer, because of the
lack of good fundamental theory for those small numbers. But
despite such uncertainty, it is still the case that there is a
rather robust conclusion that can be made: the range of sizes over
which life as we know it can exist is modest---no more than a
factor two in either direction away from the size of our universe.
The uncertainty in this estimate is considerable, but arguably
less than an order of magnitude, And the origin of the most
stringent constraint should not be surprising to those familiar
with anthropic fine-tuning arguments. It is the triple-alpha
process responsible for synthesis of carbon from helium in the
interior of stars. There is a crucial resonance in carbon, with
properties predicted in advance of its discovery. In particular
the resonance has an excitation energy which is precisely what is
needed to make the process go at the requisite rate, and which in
our context has dependence on the ratio of pion to proton
mass~\cite{ref:f,ref:o}.

\section{Implications of the Scaling Hypothesis for the Hierarchy
Problems}

If the scaling hypothesis we have made were to deliver no insight
into basic unsolved problems, it could well be argued that the
above exercise is doomed to be inconsequential speculation.
However, as we have already mentioned in the introduction, there
are interesting implications for the hierarchy problems present in
particle physics. The most quoted hierarchy problem is the
disparity between the electroweak and Planck/GUT scales. But there
are others, such as the small ratio of, say, the electron mass to
the top quark mass. And of course the small value of the
cosmological constant is another.

From the results of the previous section, these large numbers can
be ``understood". Inspection of Fig. \ref{fig:d} shows that small
universes, with sizes of order the GUT scale or smaller, do not
have much if any hierarchy problem. Small universes are
nontrivial, rather strongly coupled systems, and may well be much
more commonplace than big universes. Indeed, if one tries to
visualize the distribution in sizes of the universes comprising
the ensemble we consider, and makes the rather weak assumption
that the integral over sizes converges, then it seems natural to
make the most probable size small, rather than large. Our own
preference, based on intuition alone, is that the distribution be
approximately power-law:
\begin{equation}
R \, \frac{dN}{dR} \sim pN_{\rm
tot}\left(\frac{R_{\min}}{R}\right)^p \ .
\label{eq:7}
\end{equation}
Anthropic reasoning would suggest that the total number of
universes in this distribution which are at least as large as  our
own should be large in comparison to unity. This means that if
\begin{equation}
R_{\min} \gsim M^{-1}_{\rm planck} \ ,
\label{eq:8}
\end{equation}
then the total number of universes in the ensemble should exceed
\begin{equation}
N_{\rm tot} \sim 10^{60p} \ .
\label{eq:9}
\end{equation}
This is large, but not off scale in comparison to other numbers
which appear in cosmology, as long as $p$ has a reasonable value.
Under these circumstances, habitable universes are rare, but not
so rare as to be improbable.

Now let us return to the issue of the hierarchy problem. We have
seen, given any size distribution which is qualitatively like the
above example, that the typical, small universe has no hierarchy
problem. Our universe is atypically large, and must be so because
of the anthropic constraint that it be capable of supporting life
as we know it. And because of the scaling of parameters, this
implies that the hierarchies naturally exist in our universe.
Again, we emphasize that this argument addresses {\em all} the
hierarchies that are observed, not only the electroweak one,
provided small parameters like electron and quark masses also flow
to the Planck/GUT fixed point (not necessarily in a straight-line
manner) for universes of small size. Of course, as mentioned in
the introduction, this argument cannot really be regarded as a
definitive solution to the hierarchy problems because it is
contingent on the scaling hypothesis, which remains to be
explained. But we do see that the problem has been recast. It is
often the case in science that the recasting of a problem in
different terms can be a key to its solution. So there is good
reason to pursue further the consequences of what has been
suggested above.

The restatement of the hierarchy problem is that the rationale for
the behavior exhibited in Fig. \ref{fig:d} must be understood.
Inspection of the figure shows that (by definition), the scaling
behavior of the dark-energy scale is
\begin{equation}
\mu(R) \sim R^{-1/2}
\label{eq:10}
\end{equation}
while, to good accuracy, the behavior for the QCD scale is
\begin{equation}
\Lambda_{\rm QCD}(R) \sim R^{-1/3} \ .
\label{eq:11}
\end{equation}
The behavior for the electroweak scale is, to somewhat lesser accuracy,
\begin{equation}
v(R) \sim R^{-1/4} \ .
\label{eq:12}
\end{equation}
Therefore the recasting of the hierarchy problem reduces to the
understanding of the origin of the ``critical exponents" of 1/3
for QCD and of 1/4 for the electroweak theory. The large numbers
are gone, but mystery remains. In the next section we shall
explore a possible attack on the QCD exponent of 1/3. We have
nothing useful to say here, however, about the electroweak
1/4~\cite{ref:p}.

\section{The Critical Exponent of QCD}

The observation that the QCD scale is a factor 20/60 = 1/3 of the
way (on a logarithmic scale) between the Planck scale of
$10^{-33}$ cm and the cosmological scale of $10^{28}$ cm goes back
35 years to Zeldovich~\cite{ref:q}, and has been occasionally
rediscovered since then~\cite{ref:r}. We here approach the problem
from a perhaps surprising starting point, namely the thermodynamic
description of horizons in general relativity. The starting point
lies in work of Padmanabhan~\cite{ref:s}, who has found a simple
connection between the structure of the Einstein-Hilbert
Lagrangian and the thermodynamic description of horizons a la
Bekenstein and Hawking.

Padmanabhan begins by considering  metrics of Schwarzschild type:
\begin{equation}
ds^2 = [1-u(r)]^2 dt^2+ [1-u(r)]^{-2} dr^2 + r^2 d\theta^2 +
r^2\sin\theta\, d\phi^2
\label{eq:13}
\end{equation}
and finds that the Lagrangian $L$, as a functional of $u$, is a
perfect derivative
\begin{equation}
 L = \frac{M^2_{pl}}{16\pi} \int^r_0
d^3x\, R(u) = \frac{M^2_{pl}}{4}\int^r_0
dr\frac{\partial}{\partial r} \left[ r^2\, \frac{du}{dr} -
2ru\right] \ . \label{eq:14}
\end{equation}
If one evaluates the resulting surface integral at a horizon $R$,
where $u(R) = 1$, and where the surface gravity at the horizon
\begin{equation}
u^\prime(R) = 4\pi T \label{eq:15}
\end{equation}
can be identified with the Bekenstein-Hawking temperature $T$ as
indicated, one finds
\begin{equation}
L = \frac{1}{4} (4\pi R^2 M^2_{pl}T) - \frac{1}{2} M^2_{pl}R = T S
-  M \ . \label{eq:16}
\end{equation}
We see that the two terms can be identified as the entropy term
and energy, or mass, term respectively in the ``first law of
black-hole thermodynamics", with the correct numerical
coefficients~\cite{ref:xz}.  This can be no accident, and
undoubtedly there is much more to be understood here. Here we will
only pursue one general consequence of Padmanabhan's construction.
Let us apply it to static deSitter space, which in fact is the
spacetime appropriate to our cosmological future. In that case
\begin{equation}
u = \frac{r^2}{R^2}  = H^2 r^2 \ . \label{eq:17}
\end{equation}
We see from Eq. (\ref{eq:14}) that the Bekenstein-Hawking entropy
term, just like the accompanying energy or mass term, is cast as
an integral over a constant density. This is somewhat surprising,
since typical intuition regarding horizon entropy suggests a
visualization where the entropy is contributed by a summation over
bits of information residing on the horizon surface, of order one
bit per Planck area, rather than a summation of bits within the
bulk volume. What is instead suggested is that there are also bits
in the bulk, in one-to-one correspondence with the bits on the
surface. So it is suggestive to identify each bit residing in the
bulk with a bit residing on the surface by connecting them with a
``string". With this construction we arrive at a picture that the
objects to be counted in building up the Bekenstein-Hawking
entropy are the strings themselves, including possible structures
at the ends. In this way there is no contradiction between the two
ways of viewing the origin of the entropy. Evidently a topological
origin of this structure is suggested. The string might be a Dirac
string, or alternatively some kind of vortex structure containing
physical, as opposed to gauge, degrees of freedom. The end of the
string residing at the horizon will find itself in an ultraviolet
regime, so that its internal structure may reflect GUT symmetry
such as SU(5), while its extension into the bulk may undergo GUT
symmetry breaking, leaving a net QCD structure behind.

The connection with the QCD scaling exponent comes from estimating
the density of the entropy in the bulk. It is simply
\begin{equation}
s = \frac{S}{V} = \frac{\frac{1}{4} (4\pi R^2)M^2_{pl}}
{\frac{4}{3} \pi R^3} = \frac{3M^2_{pl}}{4R} \ .
\label{eq:18}
\end{equation}
Using
\begin{equation}
\lim_{t\rightarrow \infty} H(t) = H = 0.7 \times 10^{-10}\  {\rm
yr}^{-1}
\label{eq:18a}
\end{equation}
we obtain
\begin{equation}
s \cong (0.3\ fm^{-1})^3 \sim (\Lambda_{\rm QCD}/3)^3
\label{eq:18b}
\end{equation}
which is quite close to the observed vacuum scale of QCD. Note
that the dependence of the entropy density on the size of the
universe in Eq. (\ref{eq:18}), here represented by the radius $R =
H^{-1}$ of the deSitter horizon, does agree with the QCD scaling
exponent of 1/3. It is also interesting that the energy per bit
$\epsilon$ is simply the Bekenstein-Hawking temperature:
\begin{equation}
\epsilon = \frac{M}{N} = \frac{1}{2\pi R} = T \ . \label{eq:abc}
\end{equation}

Again, we have not solved our problems, but recast them in a
different form. It is now rephrased as an exercise in
understanding how nonperturbative structures of the QCD vacuum can
be identified with vacuum structures in cosmology, in particular
structures associated with cosmological horizons. We do not offer
any clear answer here. Our preferred hypothesis is that the
entropy is to be identified with the ``topological charge" $N$ of
the wave functional of the QCD vacuum.

Recall that the CP violating term in the QCD action
\begin{equation}
\int L\, dt = \frac{\theta}{24\pi^2} \int d^4 x E\cdot B = \theta
\int d^4x \frac{\partial K_\mu}{\partial x_\mu} = \theta \int dt
\frac{\partial N}{\partial t} \label{eq:19}
\end{equation}
is expressible in terms of the time derivative of the
``topological charge" $N$
\begin {equation}
N = \int d^3x\ K_0(x,t) \label{eq.19a}
\end{equation}
which is the quantity canonically conjugate to the $\theta$ parameter
of QCD. If $N$ indeed has a huge expectation value,
\begin{equation}
\VEV{N} \sim 10^{120}
\label{eq:20}
\end{equation}
then all the instanton-induced transitions throughout the history
of the universe will not be enough to change it significantly. To
see this assume one transition per cubic fermi per fermi of time.
It follows that
\begin{equation}
(\Delta N)^2 \sim (10^{28} \rm{cm}/10^{-13}{\rm cm})^3 \times
(10^{10}{\rm yr}/10^{-23}\sec) \sim 10^{160} \ll \VEV{N}^2 \ .
\label{eq:21}
\end{equation}
The large value of $N$ allows both it and its canonically
conjugate $\theta$ to be considered as classical variables. This in
turn may eventually shed some light on the strong CP problem. But
further investigation in this direction is beyond the scope of
this note.

In any case, the bottom line of this discussion is that the vacuum
structure of QCD may be strongly linked to the large scale
structures in cosmology, in particular horizon structures. In the
context of string-theory investigations such as AdS/CFT and its
ramifications, this may not be too surprising a conclusion.

\section{The Inflationary Universe as a Separate Universe}

We now turn to another possible ramification of the scaling idea,
one which may have too many, instead of too few experimental
consequences. We have thus far defined a universe by its size, as
determined by the value of the dark energy (or equivalently
Hubble) scale at late times. However in our own universe the
inflationary epoch contains a different, much larger dark-energy
(and Hubble) scale than at late times. Might it not therefore be
appropriate to subdivide our universe into two distinct universes?
The inflationary universe would be characterized not only by a
different dark-energy scale than the subsequent big-bang universe,
but also by different standard model parameters. Indeed,
examination of Fig.~\ref{fig:d} shows that for universes of size
$10^5$ or $10^6$ times larger than the Planck size, the QCD scale
and the electroweak scale are in the neighborhood of the GUT
scale---if anything on the high side. So a credible and simple
candidate scenario is that for this small inflationary universe
the electroweak and QCD interactions are in fact already unified,
and that the scale of their vacuum parameters might be appropriate
for identification with the inflationary dark energy scale itself!
What has to happen at the end of inflation, the so-called
``reheating" period, is in this picture quite profound---a real
change of vacuum phase, complete with a new set of standard model
parameters and with possibly a new, distinct pattern of symmetry
breaking. And as we shall mention at the end of this section, the
nature of the reheating transition from the inflationary universe
to the big-bang universe might be even more profound, involving
the basic structure of quantum theory itself.

\begin{figure}
\begin{center}
\includegraphics[width=0.8\textwidth]{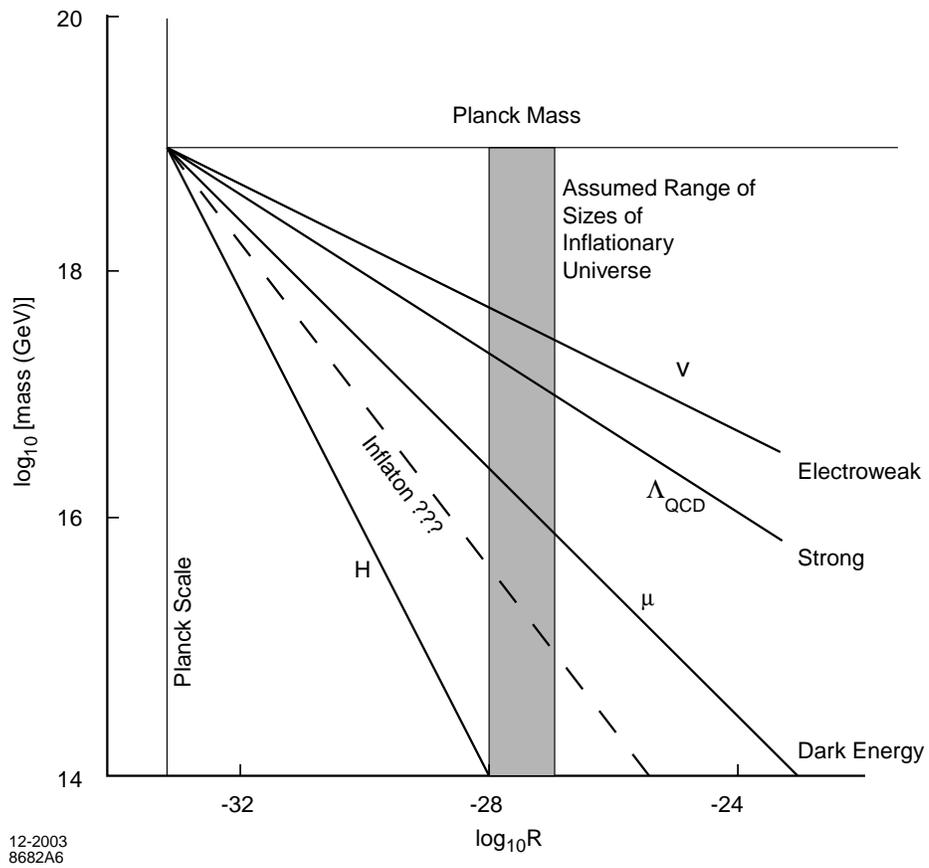}
\end{center}
\caption{Size dependence of vacuum parameters for small universes.}
\label{fig:f}
\end{figure}

\begin{figure}
\begin{center}
\includegraphics[width=0.8\textwidth]{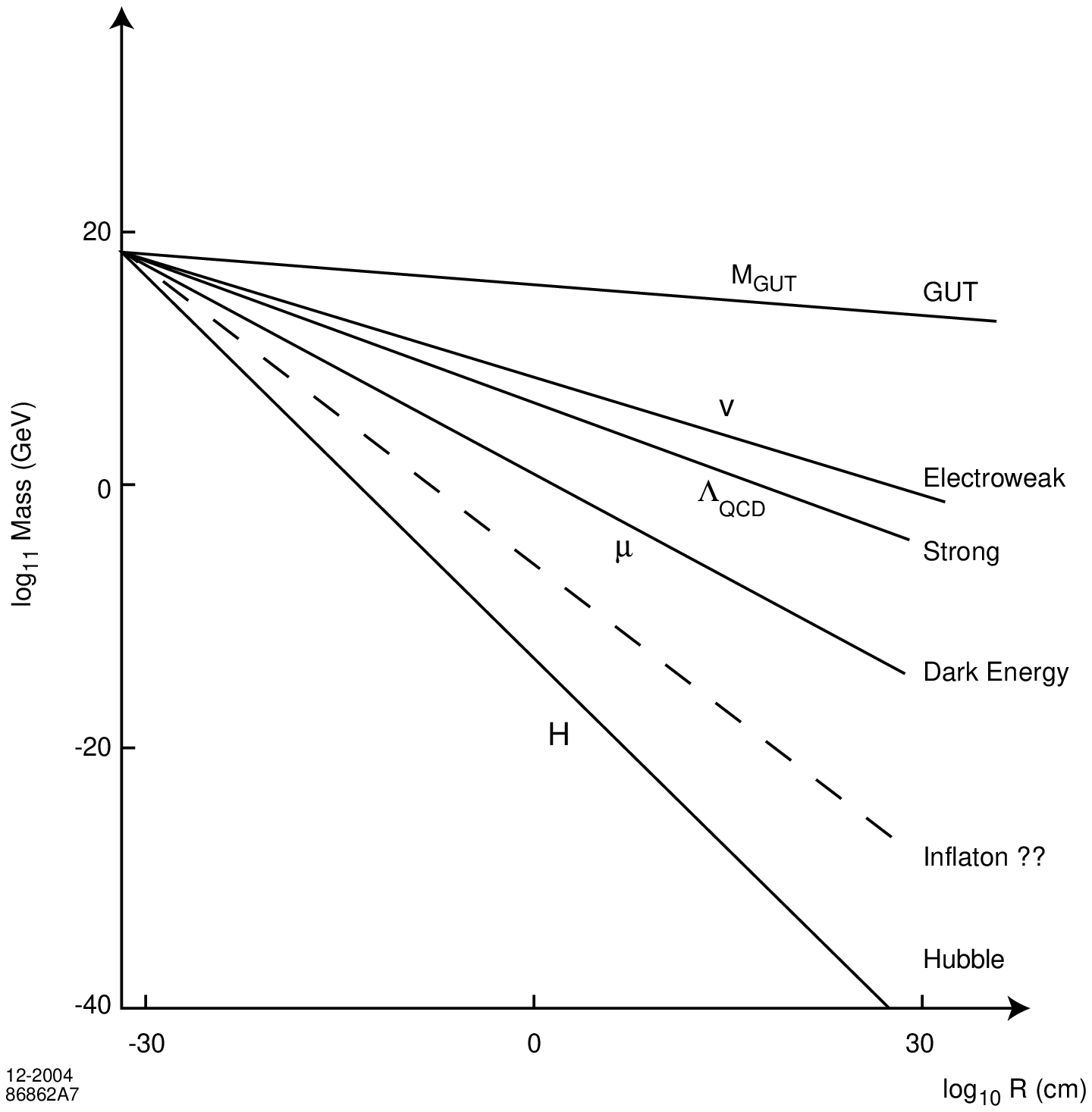}
\end{center}
\caption{Size dependence of vacuum parameters with conjectured
inflaton dark energy scale included.} \label{fig:f1}
\end{figure}

However, before considering such things, we can reverse the
procedure and consider the role of the inflaton field, both in the
small, inflationary universe, where it plays its traditional role,
and in the late, big-bang universe, where the parameter scaling
will lead to a different role. There are a large variety of
inflaton models to consider. But in this brief introductory
exploration of issues, we restrict our attention to a typical
hybrid-inflation model in order to get an idea of what is
involved. The inflaton potential is taken to be of a small scale,
much less than what accounts for the bulk of the inflationary dark
energy. The form we take is ``natural", {\em i.e.}
pseudo-Goldstone:
\begin{equation}
V(\phi) \sim m^4 \cos (\phi/F)  \ .
\label{eq:23}
\end{equation}
How this choice is fit to data on cosmic microwave background
fluctuations is a straightforward application of standard
inflation theory. The result is that the typical magnitude of the
inflaton field is of order $F$ in the above equation, with $F$
naturally taken to be of order the GUT scale. Then the magnitude
of the inflaton ``dark-energy" scale $m$, determined by the
strength of the inflaton potential, must be two to three factors
of ten smaller, in order to account correctly for the small size
of the primordial perturbations. If we then extrapolate this
potential back to our universe using the appropriate scaling law,
as shown in Fig. \ref{fig:f}, we see that the strength of the
potential for our universe is extraordinarily small---the scale of
the inflaton dark energy will be 40 or 50 powers of ten smaller
than the Planck scale (Fig.~\ref{fig:f1}). We know of no candidate
beyond-the-standard-model field which might be identified with
this inflaton.

This kind of exercise can be repeated for the whole array of
inflaton models, although no attempt will be made here to do so.
We only emphasize that there are possible observational
implications for these exercises.

Before leaving this section, we will elaborate on the comment made
above regarding the possibility that foundations of quantum
mechanics might become involved in this two-universe scenario. The
reason originates in the observation, made already in earlier
references~\cite{ref:f}, that the scaling behavior of the
couplings with size that we have described can be recast as an
overall rescaling of the standard-model Lagrangian density. In
brief, this occurs as follows. Write the Lagrangian schematically
as follows:
\begin{equation}
\mathcal{L} = F^2 + \bar\psi(\partial-eA)\psi + \phi(\partial-eA)^2
\phi + e\bar \psi \psi\phi + \lambda \phi^4 + m^2\phi^2 \ .
\label{eq:24}
\end{equation}
The terms, in order of appearance, represent (1) the gauge field
kinetic energies, (2) the quark and lepton kinetic energies, (3)
the Higgs-field kinetic energies, (4) the Higgs Yukawa couplings,
(5) the quartic Higgs self-couplings, and (6)  the Higgs mass
term. The quantity $e$ represents any of the gauge or Higgs Yukawa
couplings, all of which are assumed to have the property that
$e^{-2}$ scales logarithmically with the size of the universe, in
accordance with the behavior shown in Fig. \ref{fig:e}. The
natural assumption that this behavior will also be true for
$\lambda^{-1}$, the inverse of the quartic Higgs coupling, has
also been made. Then under the field redefinitions
\begin{equation}
eA\rightarrow A \qquad e\phi\rightarrow \phi \qquad e\psi \rightarrow \psi
\label{eq:25}
\end{equation}
we see that, except for the Higgs mass term, the dependence on
scale factors out, with an overall factor $e^{-2}$ multiplying the
Lagrangian density.  Its scale dependence is simply
\begin{equation}
\mathcal L \sim e^{-2} \mathcal L_0 \sim (\log R) \mathcal L_0 \ .
\label{eq:26}
\end{equation}
But this is tantamount to a scale dependence of the Planck
constant,
\begin{equation}
\hbar \sim (\log R)^{-1}
\label{eq:27}
\end{equation}
because its inverse is also an overall multiplier of the action.
With this interpretation, the Planck constant is (logarithmically)
large for small universes and small for large universes.

We do not know whether the above feature should be regarded as a
mere curiosity or as an indication that there is something in the
structure of quantum theory that is connected to the large-scale
geometry of the universe. Were the latter to be the case, then the
big-bang ignition, or ``reheating", transition from the
inflationary universe to the radiation-dominated universe would be
of paramount importance to fully understand. Within that dynamics
would lie some of the deepest secrets of the theory that could
possibly be revealed to us, because in that transition the value
of the Planck constant, as well as the values of the vacuum
parameters of the theory, would change discontinuously.  It would
also be imperative in that scenario to entertain the possibility
of the Planck scale and/or the speed of light also to abruptly
change during big-bang ignition.

\section{The Beginning of Inflation}

In Section 2 we used Fig. \ref{fig:b} describing the ``universe
for dummies" as evidence that its global spacetime architecture
was controlled by one large number of order $10^{30}$, the ratio
of dark energy scale to Planck scale. One of the curiosities of
that point of view is that the inflationary epoch, characterized
by large energy and small distance scales, when initiated at the
initial time we chose, is also characterized by the same number,
even though the small dark energy we observe today has not yet
entered the picture. It is arguable that this situation was
created by construction. We demanded that the starting time chosen
for evolution of our universe be identified, at least
approximately, with the time at which the size of the future event
horizon is comparable to the inverse Hubble scale. But the future
event horizon was constructed with knowledge of the properties of
the universe at future times, times when the cosmological constant
is in fact present.

Nevertheless, whatever one's point of view, there remains the
problem of how the universe during inflation, where only large
energy scales and small distances dominate, manages to evolve
during the big-bang ignition phase into a universe containing the
tiny present-day cosmological constant. It would seem that the
curiosity described in the preceding paragraph might present a
clue. One element of that clue will be the assumption that
inflation really does commence at the landmark time which we have
chosen as the starting time of our universe-in-a-box. And the idea
that we pursue is that at this initial time there is not only dark
energy present but also radiation, and perhaps ordinary matter as
well, with comparable energy densities. In other words, the
initial state of the inflationary phase is qualitatively not so
different from the state of the universe right now.  And we must
remember that the universe now is also about to enter an
inflationary phase.

\begin{figure}
\begin{center}
\includegraphics[width=0.8\textwidth]{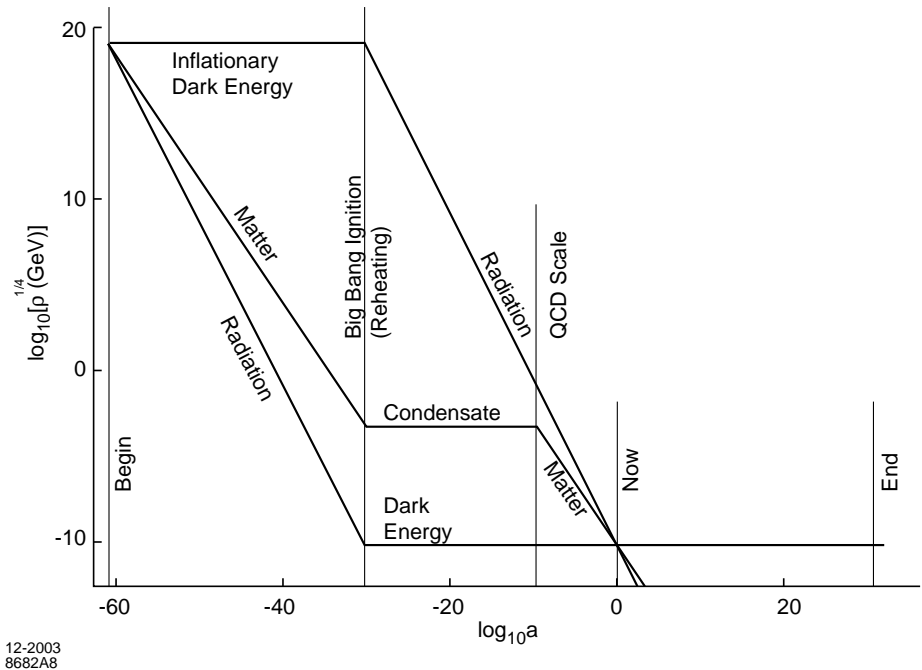}
\end{center}
\caption{Conjectured evolution of energy densities in the
``universe for dummies".} \label{fig:g}
\end{figure}

The evolution of the ``universe for dummies" from this starting
point is then straightforward: the dark energy almost immediately
dominates the Hubble parameter, and the radiation energy density
diminishes as $a^{-4}$, while the matter energy density diminishes
as $a^{-3}$. The logarithm of these energy densities are plotted
in Fig. \ref{fig:g} versus the logarithm of the scale factor. This
plot is essentially the same as Fig.~\ref{fig:d} turned on its
side, because energy densities are related to the Hubble scale in
a direct way through the FRW equation.

We see that if the radiation component we have introduced is
assumed to condense into dark energy at the end of inflation, then
the magnitude of the dark energy comes out correctly. The fate of
the matter component is also interesting. If it also condenses
into dark energy at the end of inflation, there will be far too
much of it nowadays. To account approximately for the amount of
dark matter present nowadays, this dark energy must melt into
``matter" when the Hubble parameter is 40 powers of ten larger
than the Planck scale, corresponding to the cosmic background
temperature being 20 powers of ten less than the Planck
temperature. But this is just the temperature scale to be
associated with QCD. And the behavior of this dark energy
conversion is just what happens if dark matter is composed of
axions. At the beginning of the big bang the axions behave as a
condensate, but in the neighborhood of the QCD transition the
condensate undergoes quantum fluctuations, and thereafter those
fluctuations behave as a nonrelativistic noninteracting gas in
thermal equilibrium.

\begin{figure}
\begin{center}
\includegraphics[width=0.8\textwidth]{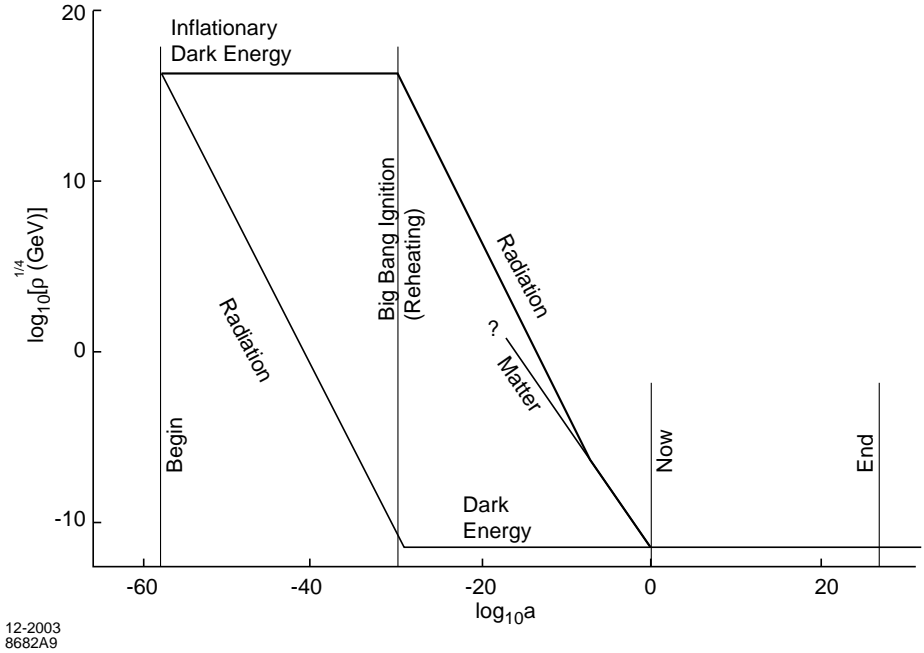}
\end{center}
\caption{Conjectured evolution of energy densities for the real
universe, without consideration of dark matter.} \label{fig:h}
\end{figure}

\begin{figure}
\begin{center}
\includegraphics[width=0.8\textwidth]{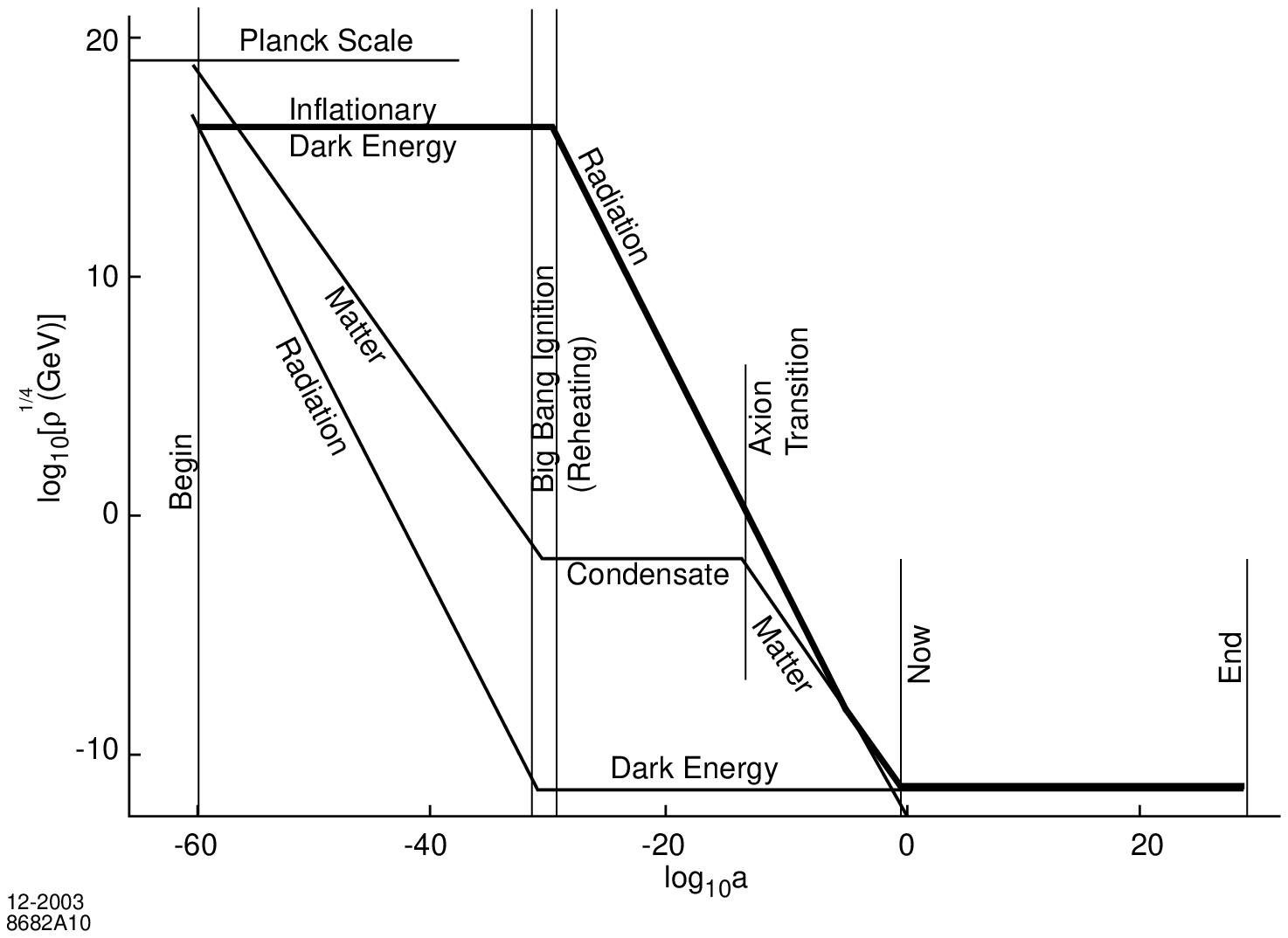}
\end{center}
\caption{Conjectured evolution of energy densities for the real
universe, with an attempted incorporation of dark energy.}
\label{fig:h1}
\end{figure}

While this scenario has a certain tidiness, and might provide a
glimpse into the prehistory of the evolution of both the dark
matter and dark energy seen nowadays, it is an idealized one,
constructed in the context of the ``universe for dummies". It then
becomes a serious and very interesting issue as to whether the
scenario can remain viable when all the real life complications
and details present in our universe are restored. This is very
much a numbers game, and a very interesting one to play. The net
result is that the scenario without the dark-matter component
appears to be quite robust (Fig.~\ref{fig:h}). But the inclusion
of the dark matter into the game is not as easy. It seems to
require an assumption that the transition from the end of
inflation to the beginning of radiation-dominated big bang is not
instantaneous. The duration of this ``reheating" epoch should
allow the scale factor $a(t)$ to increase by a factor of order a
thousand. This is illustrated in Fig. \ref{fig:h1}.

The above scenarios suggest a kind of self-similarity in the
evolution of the universe: the beginning of inflation is analogous
to the present epoch, when dark energy begins to dominate over
matter and radiation. And our future might be similar to the
inflationary epoch, perhaps terminated by another abrupt
transition to a big-bang universe containing an even smaller
cosmological constant. And there might be an epoch prior to what
we have called the beginning of inflation, again of big-bang
character, and taking one back to a truly Planckian origin. It
might also be interesting to try to combine this scenario with the
ideas of the preceding section, where each such epoch is not only
self-similar but also possesses a different set of standard model
parameters. However, the combining of the speculations in this
section with those in the previous section may not be mandatory.

\section{Concluding Remarks}

One of the main purposes of this paper has been to cast multiverse
ideology in as bottoms-up, phenomenologically driven a way as
possible. This is expedited by the definition, by construction, of
a universe which is big enough to contain essentially all of
presently conceivable phenomenology, but small enough to exclude
most of the commonplace and extravagant theoretical speculations
enveloping the subject. The experimental accuracy of the
cosmological principle, within the portion of the universe that we
have observed, then allows a very credible extrapolation to
relatively nearby regions of spacetime, where an ensemble of such
universes can be similarly constructed.

Going only this far seems to us a quite conservative procedure, at
least by contemporary standards of theoretical cosmology. More
interesting is the next step, which posits that the ensemble can
be generalized to include more distant members, of similar FRW
spacetime structure. But these members are assumed to have
different standard model parameters, in particular different
cosmological constants. In the context of present activity in
string theory, this step is also not very radical. However,
acceptance of this step then invites introduction of anthropic
reasoning, especially with respect to the role of the cosmological
constant as an anthropically determined parameter.

At this point the level of controversy increases, because as soon
as anthropic arguments are introduced, it is hard to determine
when to stop. If the cosmological constant is determined only
anthropically, what about all the other standard model parameters?
Do they not also serve as labels for the gigantic number of vacuua
in the string theory landscape? And if they are all anthropically
determined, where does that leave, say, the future program of
particle physics? The original dreams of a final theory, as
visualized two decades ago~\cite{ref:v}, lay at the opposite
extreme, with specific and naturalistic explanations of those
parameters expected to be provided by the future theory. Instead
they may be only constrained by the fact that we exist in the
universe to observe them~\cite{ref:g}.

It is not for us to dictate the answer. Maybe all the parameters
are anthropic, in which case the scientific method becomes
relatively impotent. Maybe none of them are, despite the evidence
of fine-tuning of parameters which has been uncovered and studied
by the anthropic community over the last fifty years~\cite{ref:n}.
And maybe the answer lies in between the extremes.

It is this latter hypothesis which we have chosen in this paper.
It is implemented by the presumption that, within the ensemble of
universes we have constructed, at least the principal parameters
of the standard model are strongly correlated with the value of
the cosmological constant. We have also proposed a specific form
of the correlation. This last step has been greeted  with much
skepticism, most often in a completely dismissive way.  Perhaps
this occurs because of the lack of any overarching theoretical
ideology to motivate the scaling hypothesis. In defense, we argue
from experience in searching for alternatives that the scaling
hypothesis stands out in its simplicity and
robustness~\cite{ref:yy}. Dreamers of final theories would
naturally expect most standard model parameters to be correlated
with each other, because a good theory should have very few
independent parameters. But it is {\em a priori} unlikely that the
correlations take a simple form, so simple that the right answer
can be correctly guessed. We work from a naive sense of optimism
at this point, but are rather convinced that if there {\it is} a
simple form of the correlation, the proposed correlation is likely
to be, at the least, very similar to  the correct one.

Support for this point of view comes from the output, which
provides some understanding of the hierarchy problems of particle
physics. As we discussed, the hierarchy problem is traded in for
the problem of understanding certain ``critical exponents" such as
1/3 for QCD and 1/4 for the electroweak sector. An attempt at
understanding the 1/3 of QCD was sketched in Section 6.

Here we only point out that that attempt, along with other spinoff
ideas in the subsequent two sections, are not at all abstract.
They deal with the understanding of very real issues existing
within our own universe, such as the structure of the QCD vacuum,
and the nature of the big-bang ignition (``reheating") epoch  of
our universe. It is evidence that consideration of multiple
universes may conceivably help, at a quite phenomenological level,
in uncovering the nature of physical processes in our own
universe.

\section*{Acknowledgment}

We thank M. Weinstein and R. Akhoury for valuable discussions and
criticism.  Some of this work was prepared for the workshop
``Universe or Multiverse?", sponsored by the John Templeton
Foundation.  The author thanks the Foundation for the invitation
to that meeting and financial support.

\setcounter{equation}{0}

\section*{Appendix A: Cosmology 101}

We here record equations used in constructing the figures in the
text.  We do assume familiarity with the physics of big-bang
cosmology, as described in standard texts.

The FRW equation is
\begin{equation}
H^2(t) = \frac{8\pi}{3M^2_{pl}}\, \rho(t) \label{eq:A1}
\end{equation}
with $\rho(t)$ the energy density of sources, and the Hubble
expansion parameter
\begin{equation}
H(t) = \frac{\dot a(t)}{a(t)} \ .
\label{eq:A2}
\end{equation}
The equation of state of the source of energy must be specified.
Generically, one writes
\begin{equation}
p = \omega \rho
\label{eq:A3}
\end{equation}
with $\omega = 0,\ \frac{1}{3},$ and $-1$ for matter, radiation,
and dark energy, respectively.  Then energy-momentum conservation demands
\begin{equation}
d(\rho a^3) = - p\, d(a^3) \ .
\label{eq:A4}
\end{equation}
Solution of these equations for the scale factor yields
\begin{equation}
a(t) \sim \left\{
\begin{array}{l@{\qquad}l}
              t^{1/2} & {\rm Radiation} \\
              t^{2/3} & {\rm Matter} \\
               e^{Ht} & {\rm Dark\  energy}
\end{array} \right.
\label{eq:A5}
\end{equation}
The behavior of $H$ versus $a$ is then
\begin{equation}
H(a) \sim \left\{
\begin{array}{l@{\qquad}l}
a^{-2} & {\rm Radiation} \\
a^{-3/2} & {\rm Matter} \\
{\rm constant} & {\rm Dark\ energy}
\end{array} \right. \ .
\label{eq:A6}
\end{equation}
These results are readily checked by substitution in Eqs.
(\ref{eq:A1}) and (\ref{eq:A2}).  In Fig.~\ref{fig:d} we assume
abrupt transitions between the various epochs shown.  At the
transitions both $H$ and $a$ must be continuous. From the equation
of state for radiation we readily deduce that the mean wavenumber
$k$ of photons in the cosmic microwave background scales as
\begin{equation}
k  \sim a^{-1} \ , \label{eq:A8}
\end{equation}
whether or not the photons are in thermal equilibrium.  When thermal
equilibrium applies, evidently $k$ is a measure of temperature.

One important simplification we have made occurs in the
radiation-dominated epoch.  The equation of state is, expressed in
terms of temperature,
\begin{equation}
\rho = g(T)\, T^4
\label{eq:A9}
\end{equation}
and the density of states factor $g(T)$ decreases by more than an
order of magnitude during expansion.  This changes slightly the
plotted curves.  But on the scale shown, it is scarcely
noticeable.

In the deSitter epoch in our future, Eq. (\ref{eq:6}) for the
future event horizon simplifies to
\begin{equation}
\label{eq:A10} E(t) = H^{-1} \ ,
\end{equation}
while in the matter and radiation-dominated eras,
\begin{equation}
E(t) \approx a(t) \int^\infty_0 \frac{dt^\prime}{a(t^\prime)} =
a(t)\, \eta_0\ . \label{eq:A11}
\end{equation}
In that epoch, the slope of the curve for log $E(t)$ versus
log$\,a$ is $+1$, as shown in Figs. \ref{fig:b} and \ref{fig:c}.

\setcounter{equation}{0}



\begin{thebibliography}{99}
\bibitem{ref:d}
This conclusion may still be reached by a modern family, with
small children, while driving across that great state.
\bibitem{ref:xz}
We note that the relation $M=\frac{1}{2}\, M^2_{pl}R$ is true for
both a Schwarzschild black hole or for the total dark energy
inside a static deSitter horizon.
\end{thebibliography}
\end{document}